\input{aipcheck}

\documentclass[final
    ,final            
  ]
  {aipproc}

\layoutstyle{6x9}

\def\littleprime{\ifmmode{\scriptscriptstyle \prime }
    \else{\hbox{$\scriptscriptstyle \prime$ }}\fi}
\def\littlecirc{\ifmmode{\scriptscriptstyle \circ }
    \else{\hbox{$\scriptscriptstyle \circ $ }}\fi}
\def\littless{\ifmmode{\scriptscriptstyle s }
    \else{\hbox{$\scriptscriptstyle s $ }}\fi}
\def\arcsec{\raise .9ex \hbox{\littleprime\hskip-3pt\littleprime}}
\def\arcmin{\raise .9ex \hbox{\littleprime}}
\def\degree{\raise .9ex \hbox{\littlecirc}}
\def\arcsecpoint{\hbox to 1pt{}\rlap{\arcsec}.\hbox to 2pt{}}
\def\arcminpoint{\hbox to 1pt{}\rlap{\arcmin}.\hbox to 2pt{}}
\def\degreepoint{\hbox to 1pt{}\rlap{\degree}.\hbox to 2pt{}}
\def\gtapr {\lower .1ex\hbox{\rlap{\raise .6ex\hbox{\hskip .3ex
        {\ifmmode{\scriptscriptstyle >}\else
                {$\scriptscriptstyle >$}\fi}}}
        \kern -.4ex{\ifmmode{\scriptscriptstyle \sim}\else
                {$\scriptscriptstyle\sim$}\fi}}}
\def\ltapr {\lower .1ex\hbox{\rlap{\raise .6ex\hbox{\hskip .3ex
        {\ifmmode{\scriptscriptstyle <}\else    
                {$\scriptscriptstyle <$}\fi}}}
        \kern -.4ex{\ifmmode{\scriptscriptstyle \sim}\else
                {$\scriptscriptstyle\sim$}\fi}}}

\begin{document}

\title{Linear radio structures in selected Seyfert and LINER galaxies}


\author{E. Xanthopoulos}{
  address={University of California Davis, Department of Physics, Davis, CA 95616}
  ,altaddress={IGPP/Lawrence Livermore National Laboratory, Livermore, CA 94550 }
}

\author{A. H. Thean }{
  address={University of Manchester, Jodrell Bank Observatory, Macclesfield, Cheshire SK11 9DL, England}
}

\author{A. Pedlar }{
  address={University of Manchester, Jodrell Bank Observatory, Macclesfield, Cheshire SK11 9DL, England}
}

\author{A. M. S. Richards }{
  address={University of Manchester, Jodrell Bank Observatory, Macclesfield, Cheshire SK11 9DL, England}
}

\begin{abstract}
High resolution MERLIN 5 GHz observations (0\arcsecpoint04) of 7
Seyfert galaxies, selected as the ones previously showing evidence
of collimated ejection, have been compared with high resolution
archive HST data. The radio maps reveal rich structures in all
the galaxies. NGC 2639 and TXFS 2226-184 have multiple knot parsec-scale
extended structures, Mrk 1034, Mrk 1210, NGC 4922C and NGC 5506
reveal one-sided jets, while IC 1481 exhibits a jet-like
extension. The close correlation between the radio-emitting
relativistic plasma and the ionized gas in the inner regions of
these galaxies allows us to study in detail the physics close to
the center of low luminosity AGN.

\end{abstract}

\maketitle


\section{Introduction}

High resolution radio observations have revealed double and triple
radio sources in many Seyfert nuclei, suggesting that the
collimated ejection of material is occurring in these objects in
the same way as the radio jets in quasars and radio galaxies. But
very few

\begin{figure}[h]
\begin{minipage}[t]{7.5cm}
\includegraphics[width=0.93\textwidth]{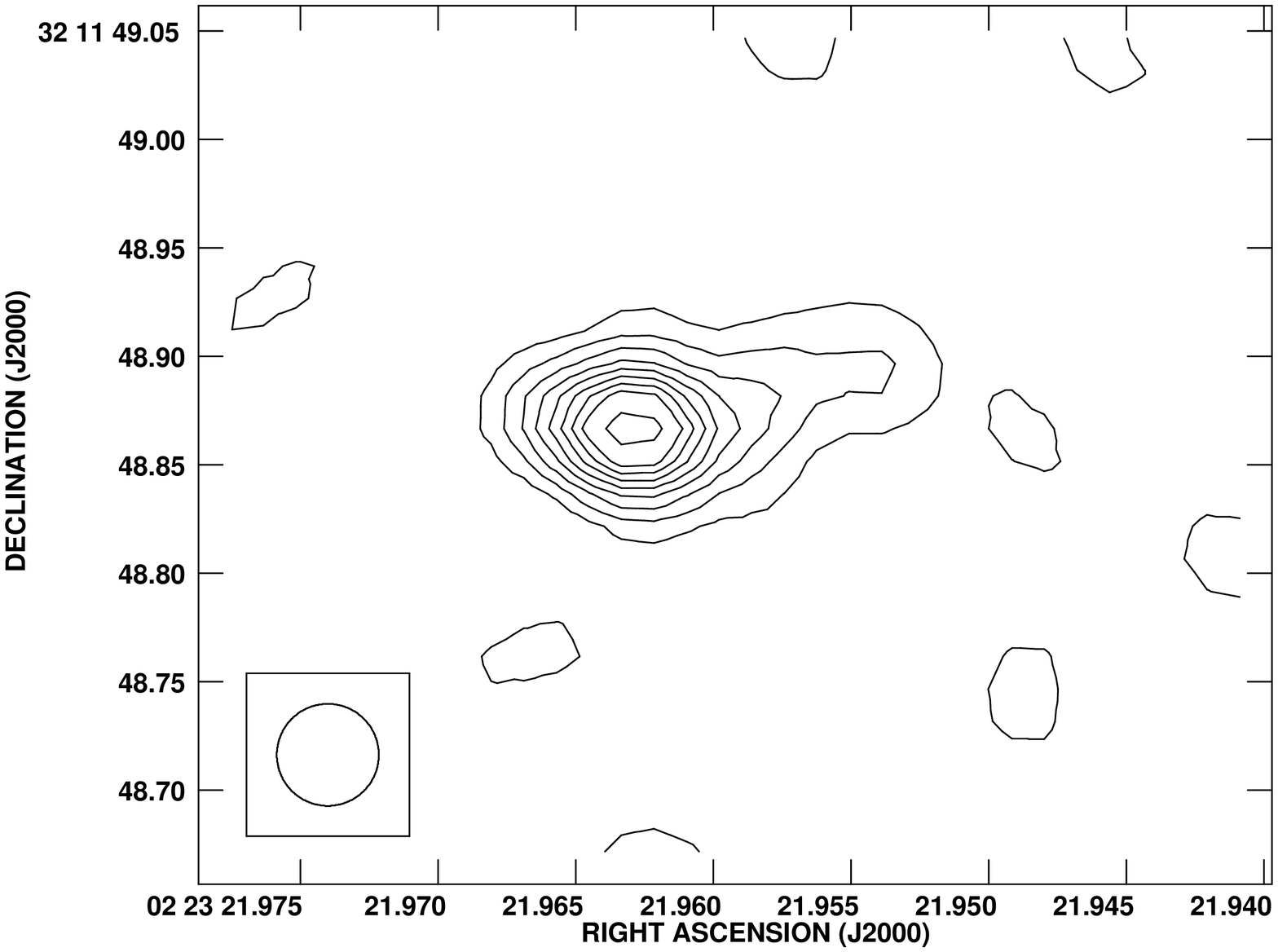}
\end{minipage}
\hfill
\begin{minipage}[t]{7.5cm}
\includegraphics[width=0.77\textwidth]{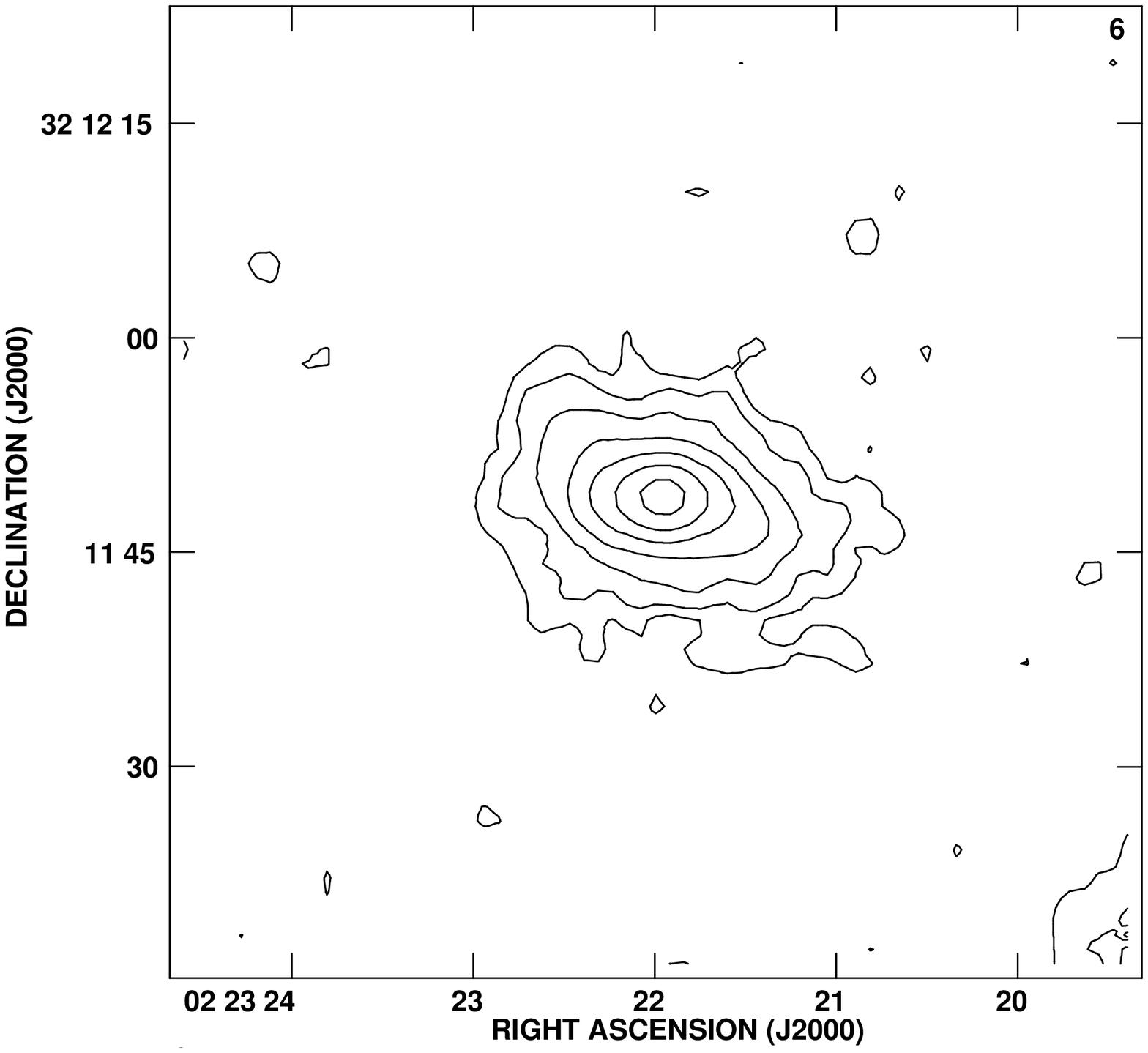}
\caption{(Left:) 5 GHz MERLIN map of Mrk 1034. Contour levels are at
1.978e-4$\times$(1,2,3,4,5,6,7,8,10,16,32,64) Jy/beam. The peak
flux is 2.219 mJy/beam and the rms noise level is$\sim$98
$\mu$Jy/beam. (Right:) K$_{s}$ (2.17 $\mu$m) Two Micron All Sky
Survey (2MASS) image of Mrk 1034.} \label{mkn1034}
\end{minipage}
\hfill
\end{figure}

\noindent Seyfert galaxies have been found to have radio
structures that can be described as jets (Kukula et al. 1993;
Ghosh et al. 1994). The angular resolution of MERLIN at 5 GHz is
equivalent to that of the {\it Hubble Space Telescope} (HST),
making these radio images ideal for comparison with the structure
of the narrow-line region (NLR) and extended narrow line region
(ENLR) and perfect cases to study in detail the individual small
scale ``jets'' as revealed in e.g. Capetti et al. (1999). The
purpose of the present study is to: {\bf a)} Investigate the
variety of collimated ejection in low luminosity AGN and increase
the small number of multiple component ``jets'' known, {\bf b)}
Determine the alignment of the collimated ejection on scales of a
few parsecs and compare this alignment with estimates made from
extended NLR studies from HST and other optical observations, {\bf
c)} Involve hydrodynamical simulations and multi-wavelength data
in order to constrain the bow shock and similar models and get a
more complete picture of this class of AGN and, {\bf d)} Compare
the 5 GHz continuum observations with 22 GHz MERLIN observations
of the same galaxies, since the coincidence of the position of the
two emissions may mark the location of the obscured central
engine.

\section{Maps and Results}

\noindent {\bf Mrk 1034:} The MERLIN 5 GHz map
(Fig.~\ref{mkn1034}) of Ark 81 - one of the Seyfert 1 galaxies of
the pair Ark 80, Ark 81 that are interconnected- shows an E-W
elongated radio structure (20 mas) with a 10 mas jet extension to
the E consistent with a highly collimated radio jet in that
direction. Mrk 1034 has not been observed yet with HST. However,
the inner contours of a 2MASS near-IR image of Mrk 1034
(Fig.~\ref{mkn1034}) follow the same elongation along the E-W
direction.

\begin{figure}[h]
\begin{minipage}[t]{7.5cm}
\includegraphics[width=0.88\textwidth]{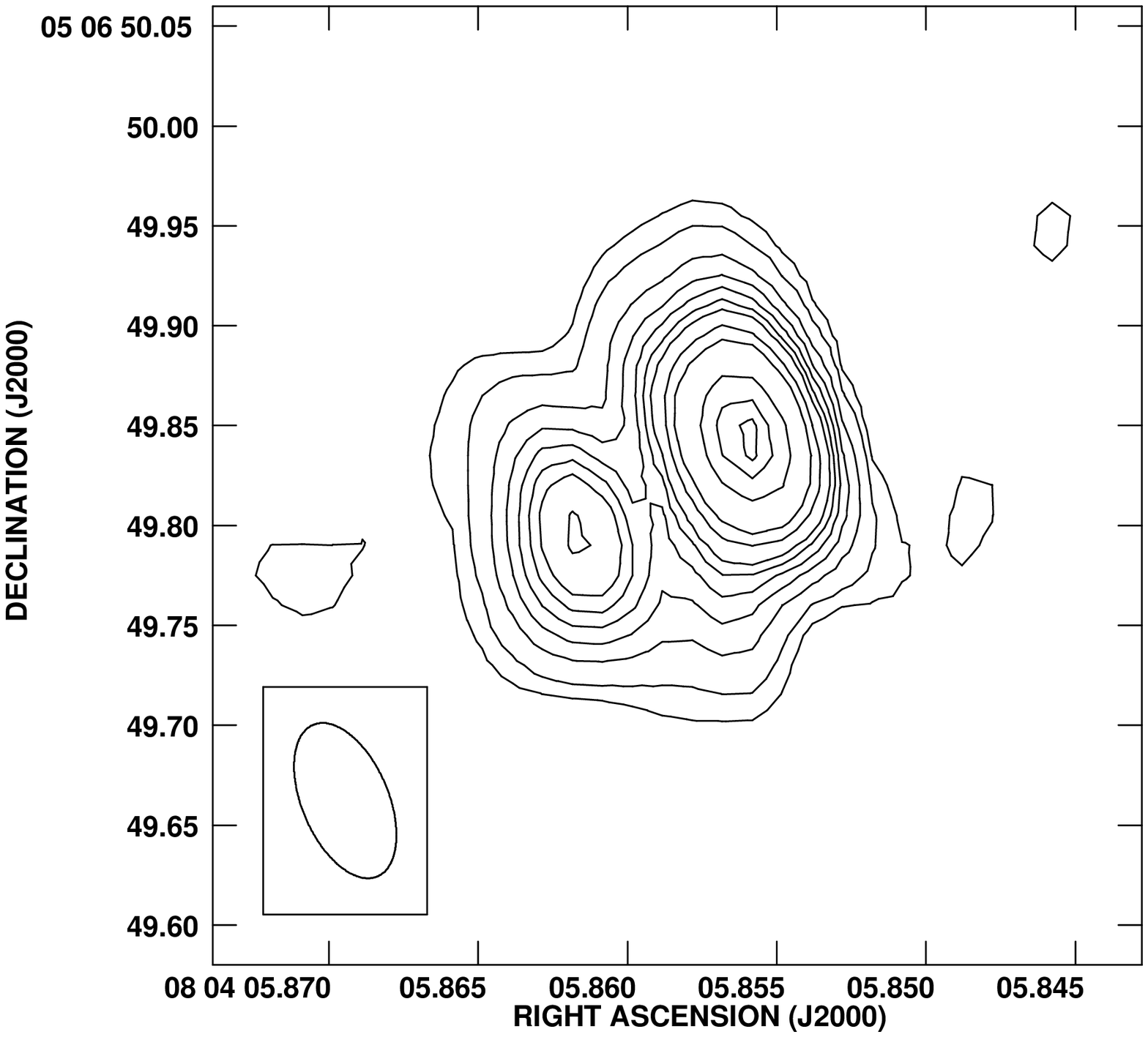}
\end{minipage}
\hfill
\begin{minipage}[t]{7.5cm}
\includegraphics[width=1.0\textwidth]{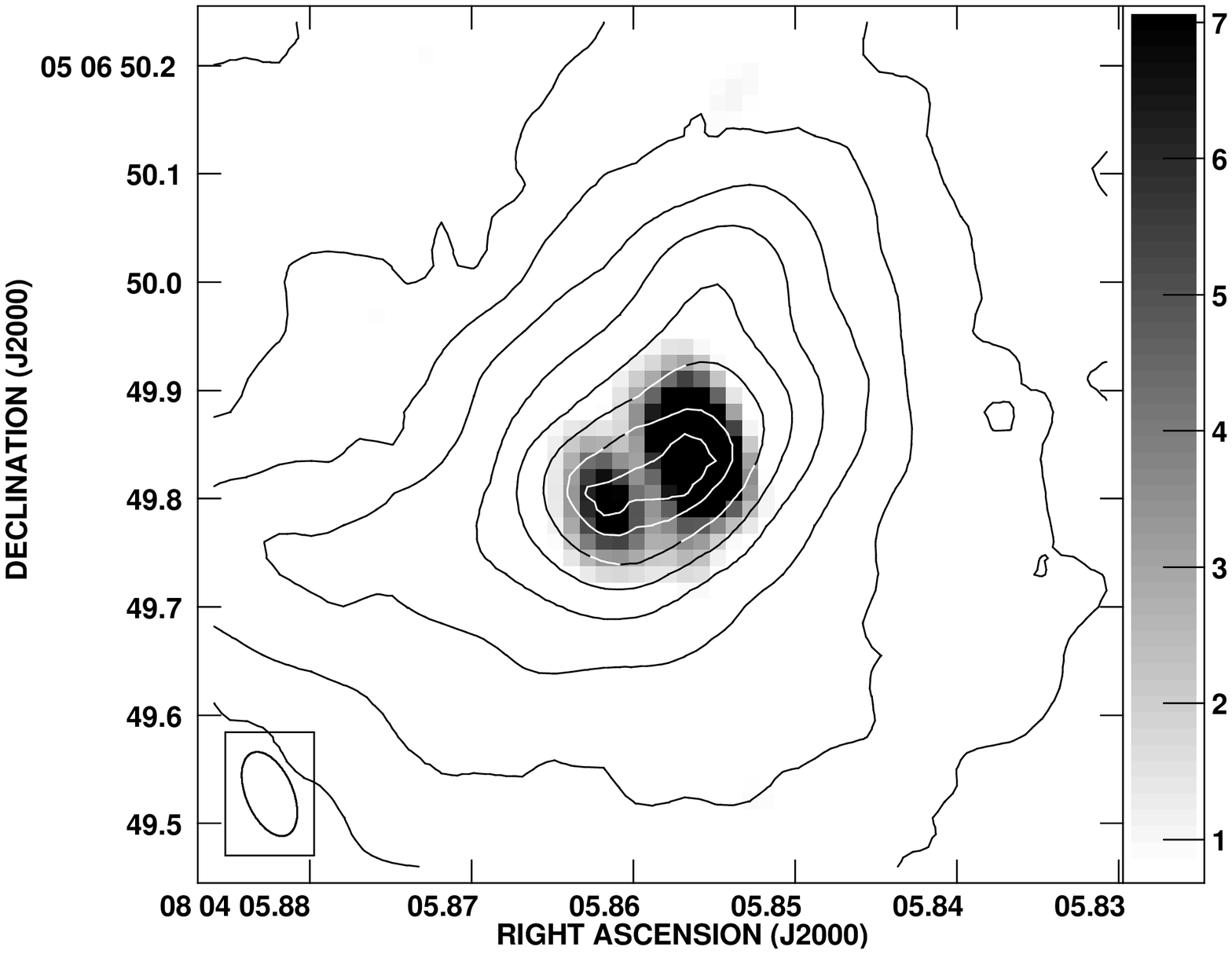}
\caption {(Left:) 5 GHz MERLIN map of Mrk 1210. Contour levels are
at 4.899e-4$\times$(1,2,4,6,8,10,12,16,20,32,40,45) Jy/beam. The
peak flux is 22.83 mJy/beam and the rms noise level is $\sim$ 163
$\mu$Jy/beam. (Right:) HST ACS/HRC (F330W) contour map of Mrk 1210
overlaid on the MERLIN 5 GHz greyscale image.} \label{mkn1210}
\end{minipage}
\hfill
\end{figure}

\vspace{0.15cm} \noindent {\bf Mrk 1210:} The near-UV emission
revealed in a high resolution (0\arcsecpoint027) ACS/HRC HST
(F330W) image of Mrk 1210 follows closely the $\sim$15 mas
one-sided jet unveiled in the MERLIN 5 GHz map
(Fig.~\ref{mkn1210}) of this amorphous Seyfert 2 galaxy.

\begin{figure}[h]
\begin{minipage}[t]{7.5cm}
\includegraphics[width=1.0\textwidth]{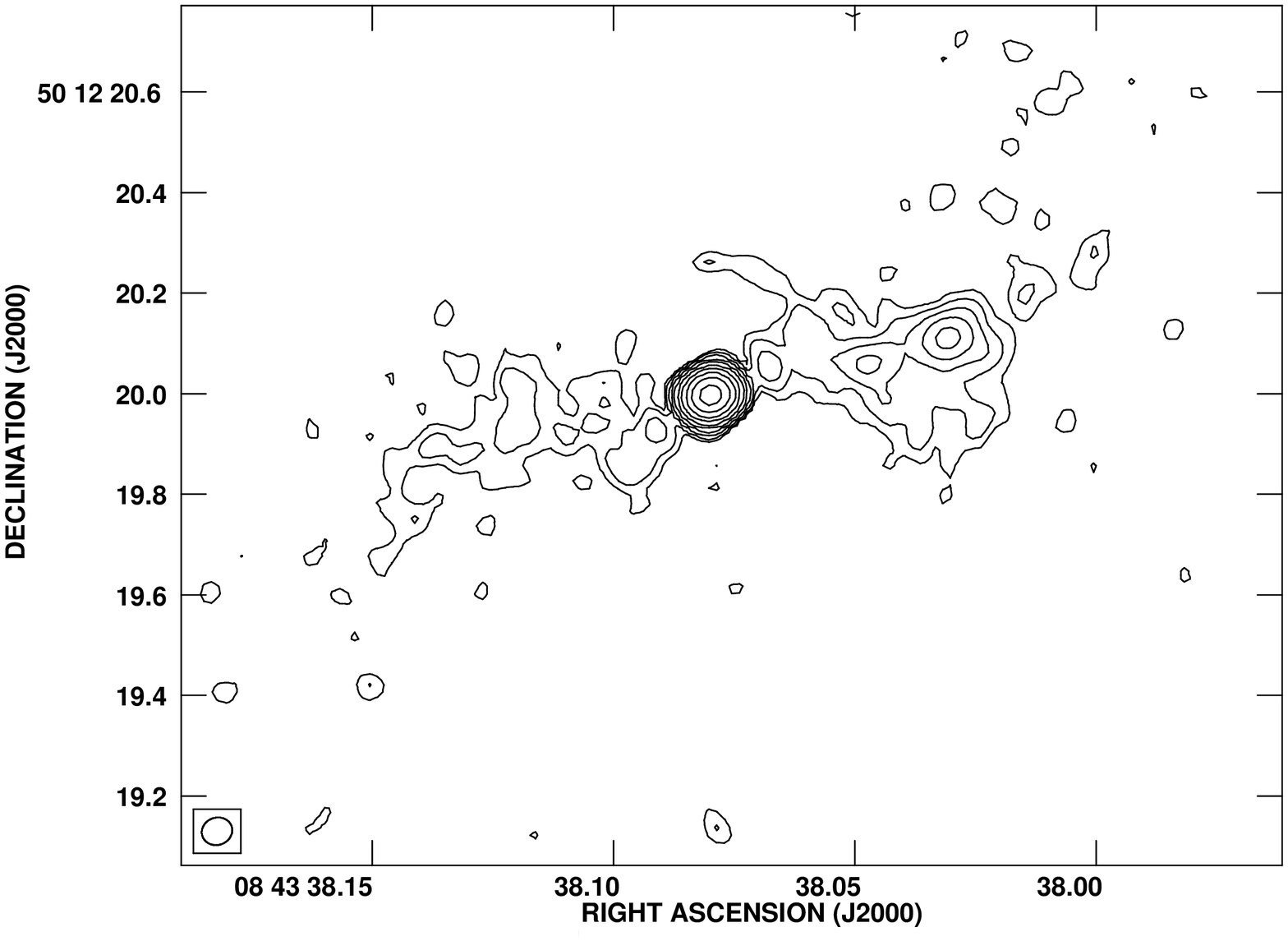}
\end{minipage}
\hfill
\begin{minipage}[t]{7.5cm}
\includegraphics[width=0.88\textwidth]{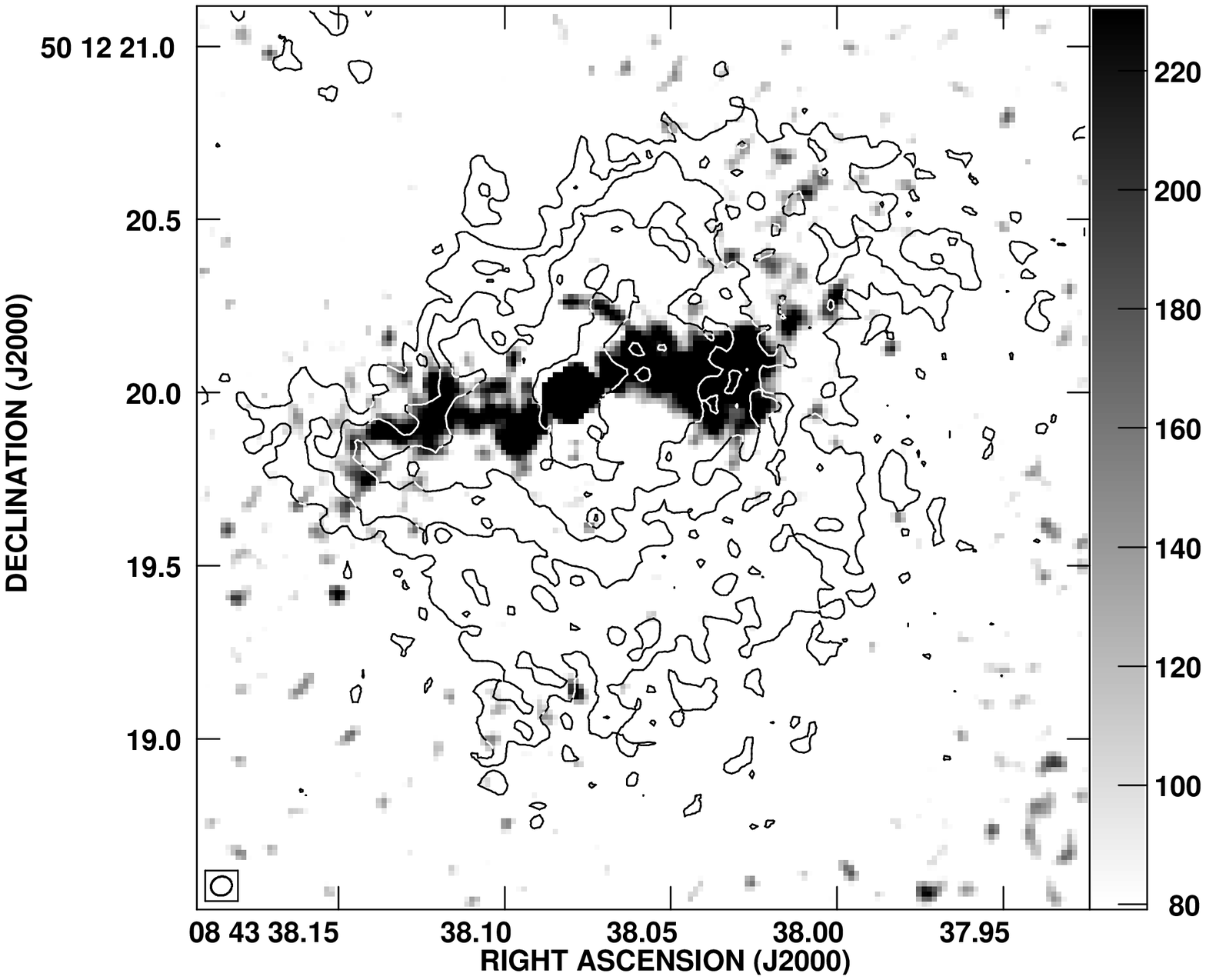}
\caption {(Left:) 5 GHz MERLIN map of NGC 2639. Contour levels are
at 1.297e-4$\times$(1,2,4,8,16,32,64,128,256,512) Jy/beam. The
peak flux is 93.5 mJy/beam and the rms noise level is $\sim$48
$\mu$Jy/beam. (Right:) HST ACS/HRC (F330W) contour map of NGC 2639
overlaid on the MERLIN 5 GHz greyscale image.} \label{ngc2639}
\end{minipage}
\hfill
\end{figure}

\vspace{0.15cm} \noindent {\bf NGC2639:} The MERLIN 5 GHz map
(Fig.~\ref{ngc2639}) of this well-known LINER megamaser galaxy
reveals a bright core and symmetrical E-W ``wings'' with rich
structure (multiple knots), that is highly correlated with the UV
star formation morphology of the HST ACS/HRC image
(Fig.~\ref{ngc2639}).

\begin{figure}[h]
\begin{minipage}[t]{7.5cm}
\includegraphics[width=0.85\textwidth]{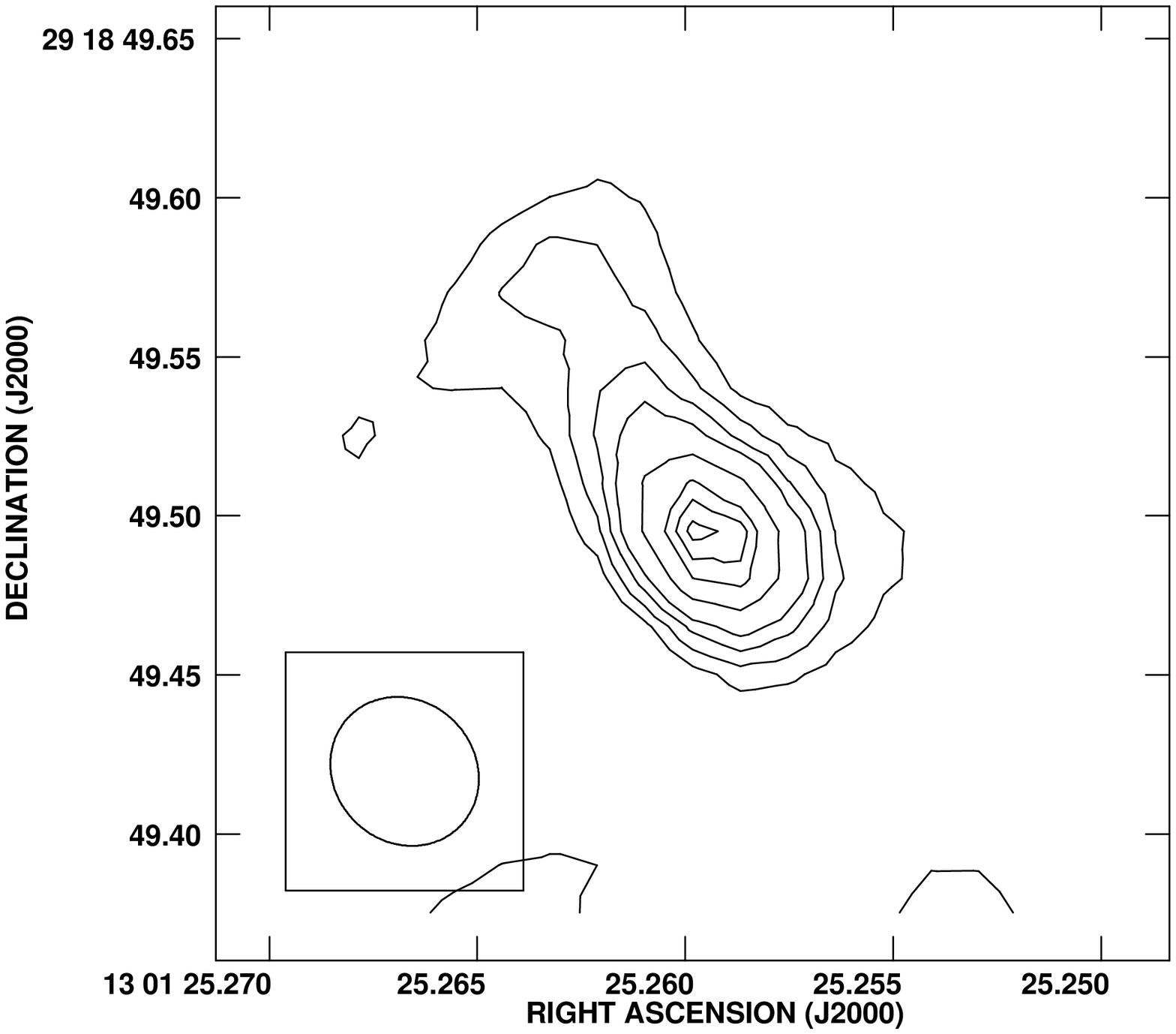}
\end{minipage}
\hfill
\begin{minipage}[t]{7.5cm}
\includegraphics[width=0.7\textwidth]{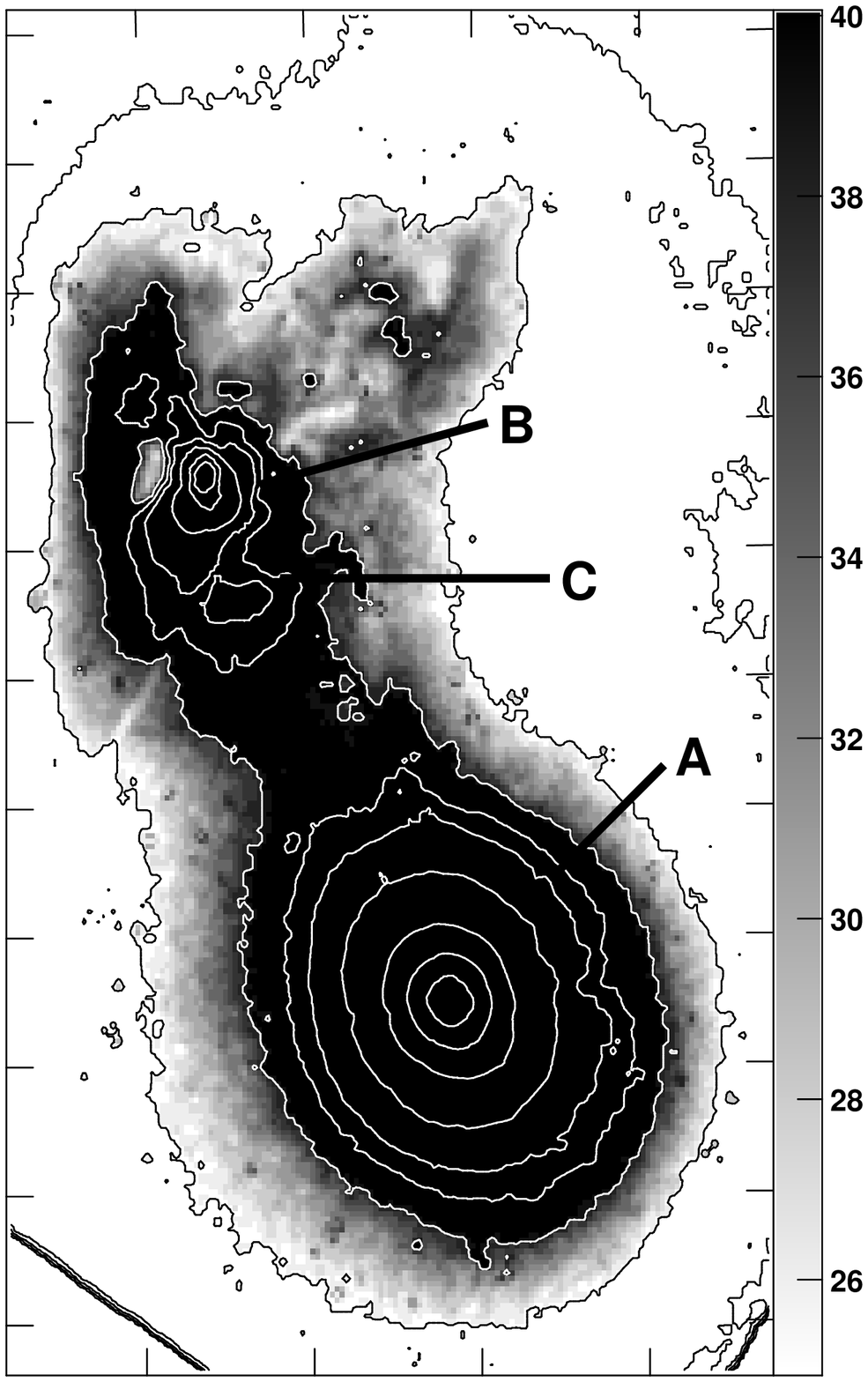}
\caption {(Left:) 5 GHz MERLIN map of NGC 4922C (PGC~044896).
Contour levels are at
3.803e-4$\times$(1,2,4,6,8,9,10,16,32,64,128,256,512,1024)
Jy/beam. The peak flux is 3.97 mJy/beam and the rms noise level is
$\sim$126 $\mu$Jy/beam. (Right:) HST WFPC2 (F606W) image of the
triple system NGC 4922.} \label{ngc4922}
\end{minipage}
\hfill
\end{figure}

\vspace{0.15cm} \noindent {\bf NGC 4922:} The MERLIN 5 GHz map
(Fig.~\ref{ngc4922}) detects the middle component C, PGC~044896
(FIRST J130125.2+291849 radio source), from the three (3) galaxy
system NGC 4922 (NGC 4922A/B are both Seyfert 2) as shown in HST
WFPC2 (F606W) image (Fig.~\ref{ngc4922}). The radio structure is
elongated in the NE-SW direction, in good agreement with the
orientation of the interacting galaxies.

\begin{figure}[h]
\begin{minipage}[t]{7.3cm}
\includegraphics[width=0.744\textwidth]{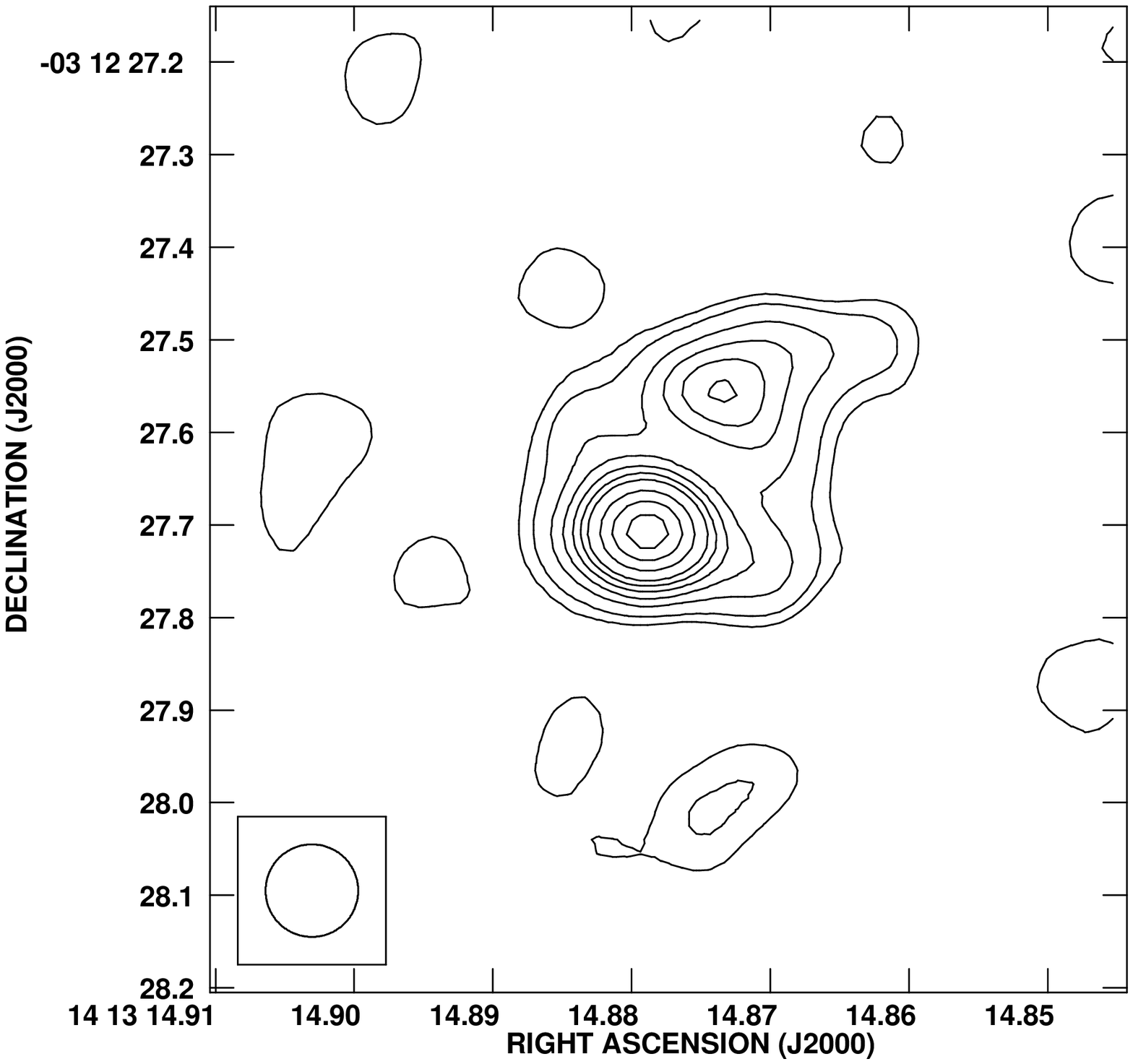}
\end{minipage}
\hfill
\begin{minipage}[t]{7.5cm}
\includegraphics[width=1.0\textwidth]{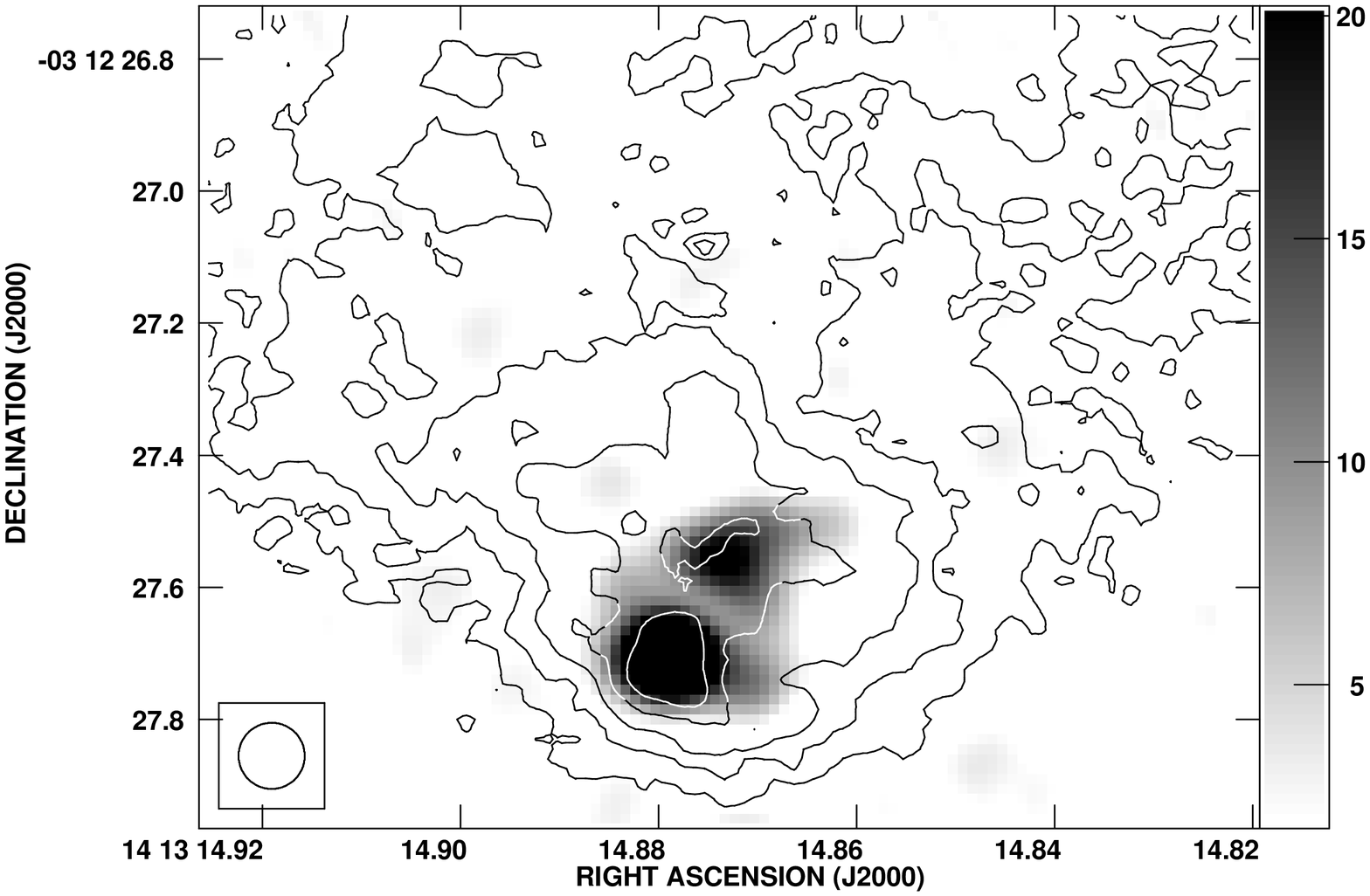}
\caption {(Left:) 5 GHz MERLIN map of NGC 5506. Contour levels are
at 2.185e-3$\times$(1,2,4,6,10,12,16,20,24) Jy/beam. The peak flux
is 57.29 mJy/beam and the rms noise level is $\sim$ 546
$\mu$Jy/beam. (Right:) HST ACS/HRC (F330W) contour map of NGC 5506
overlaid on the MERLIN 5 GHz greyscale image. } \label{ngc5506}
\end{minipage}
\hfill
\end{figure}

\vspace{0.15cm} \noindent {\bf NGC 5506:} The MERLIN 5 GHz map
shows a NW radio loop/jet ($\sim$5 mas in extent) that follows
closely the rich near-UV emission (Fig.~\ref{ngc5506}).

\begin{figure}[h]
\begin{minipage}[t]{7.5cm}
\includegraphics[width=0.8\textwidth]{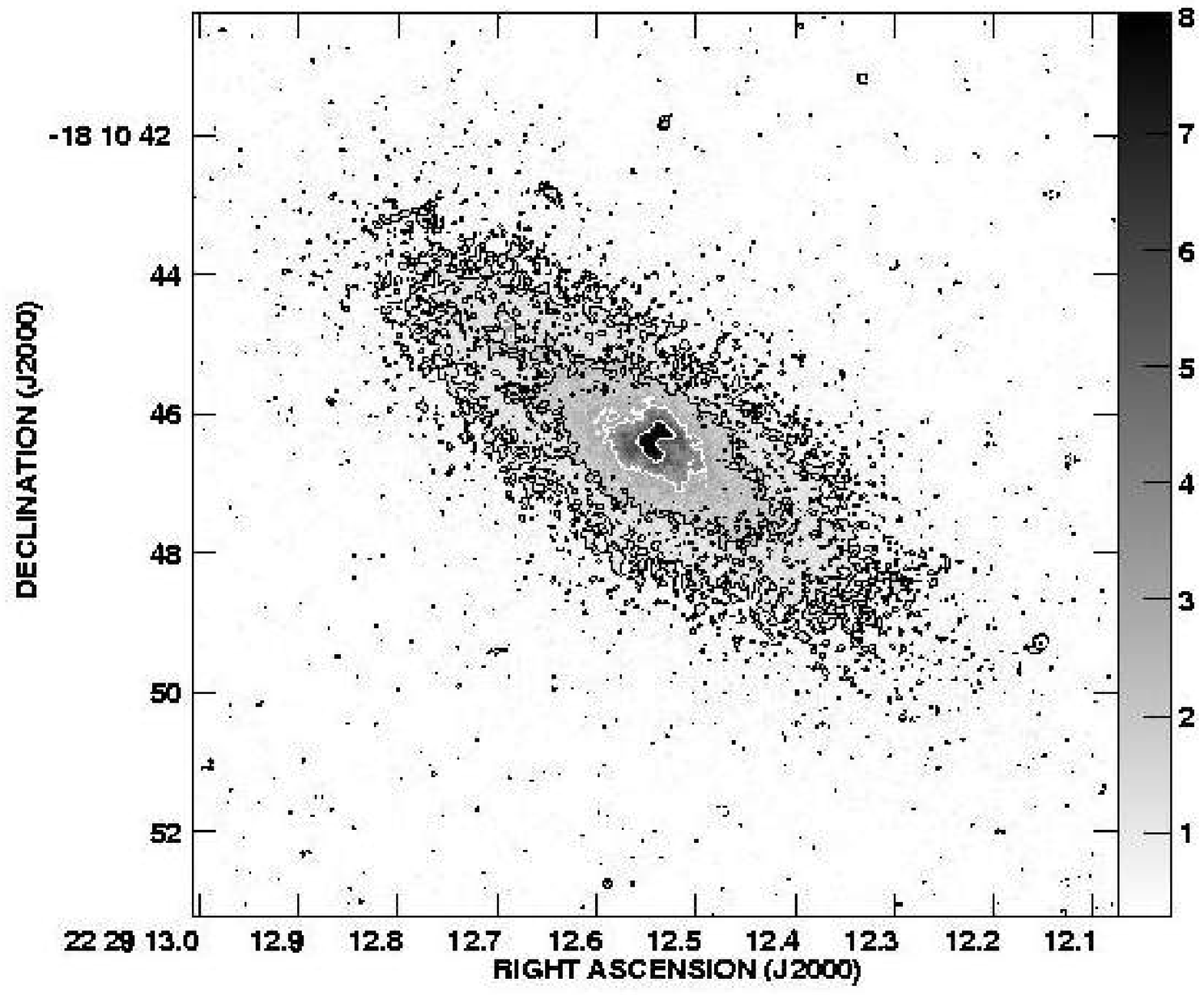}
\end{minipage}
\hfill
\begin{minipage}[t]{7.35cm}
\includegraphics[width=0.8\textwidth]{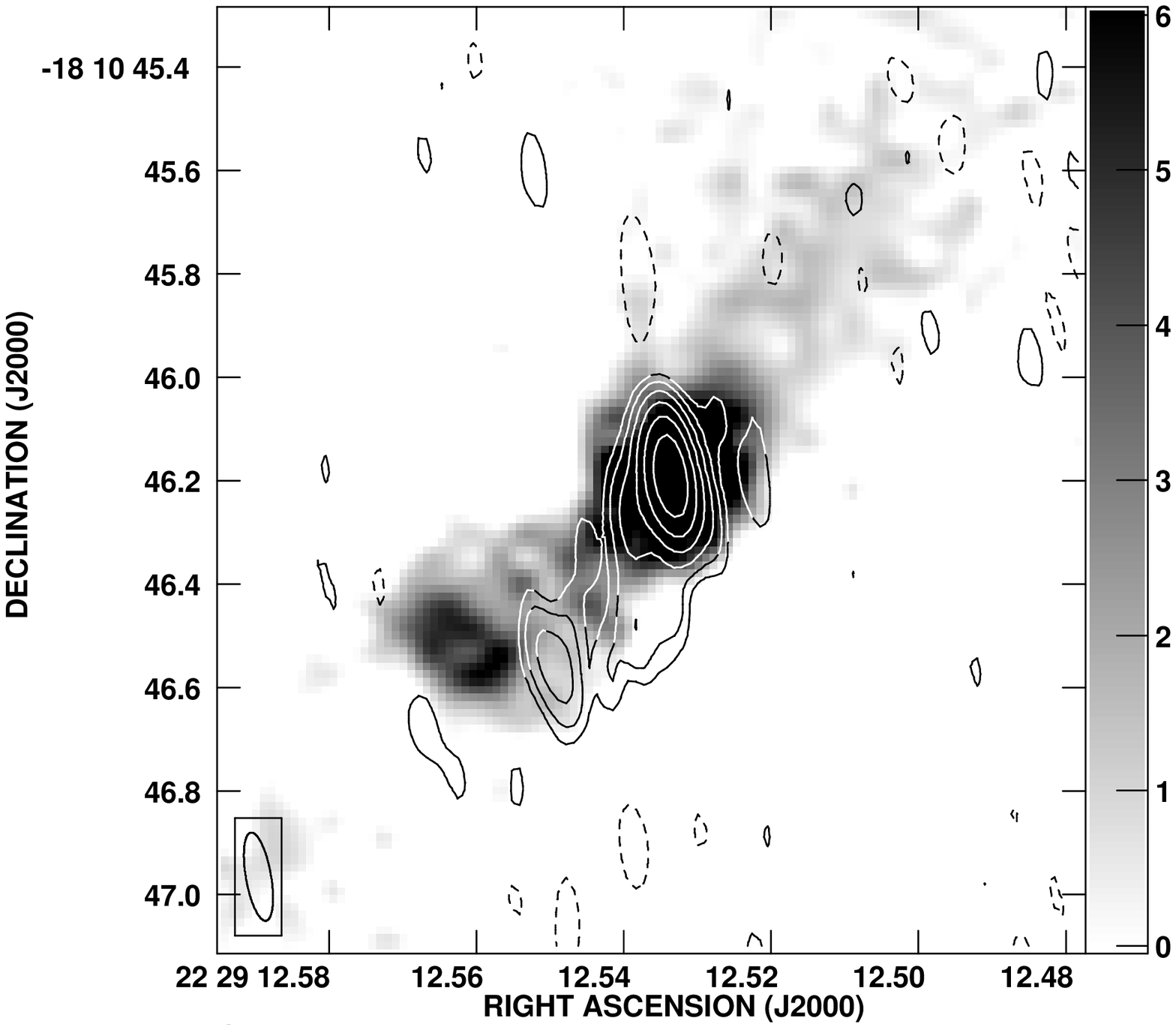}
\caption {(Left:) The average red/green continuum HST WFPC2 image
of the host galaxy of TXFS 2226-184. (Right:) 5 GHz MERLIN map of
TXFS 2226-184 overlaid on the HST WFPC2 H$\alpha$ image. Contour
levels are at 1.750e-4$\times$(1,2,4,8,16,32,64,128,256,512)
Jy/beam. The peak flux is 14.25 mJy/beam and the rms noise level
is $\sim$ 65 $\mu$Jy/beam.} \label{txfs2226}
\end{minipage}
\hfill
\end{figure}

\vspace{0.15cm} \noindent {\bf TXFS 2226-184:} The multiple knot
radio structure (Fig.~\ref{txfs2226}) coincides with the nucleus
of the host galaxy and then is aligned with the H$\alpha$ emission
on the larger scale. This alignment reflects an interaction
between the radio jet and the interstellar medium. The H$\alpha$
and radio structures are perpendicular to the galaxy major axis
(Fig.~\ref{txfs2226}).

\begin{figure}[h]
\begin{minipage}[t]{4.7cm}
\includegraphics[width=0.94\textwidth]{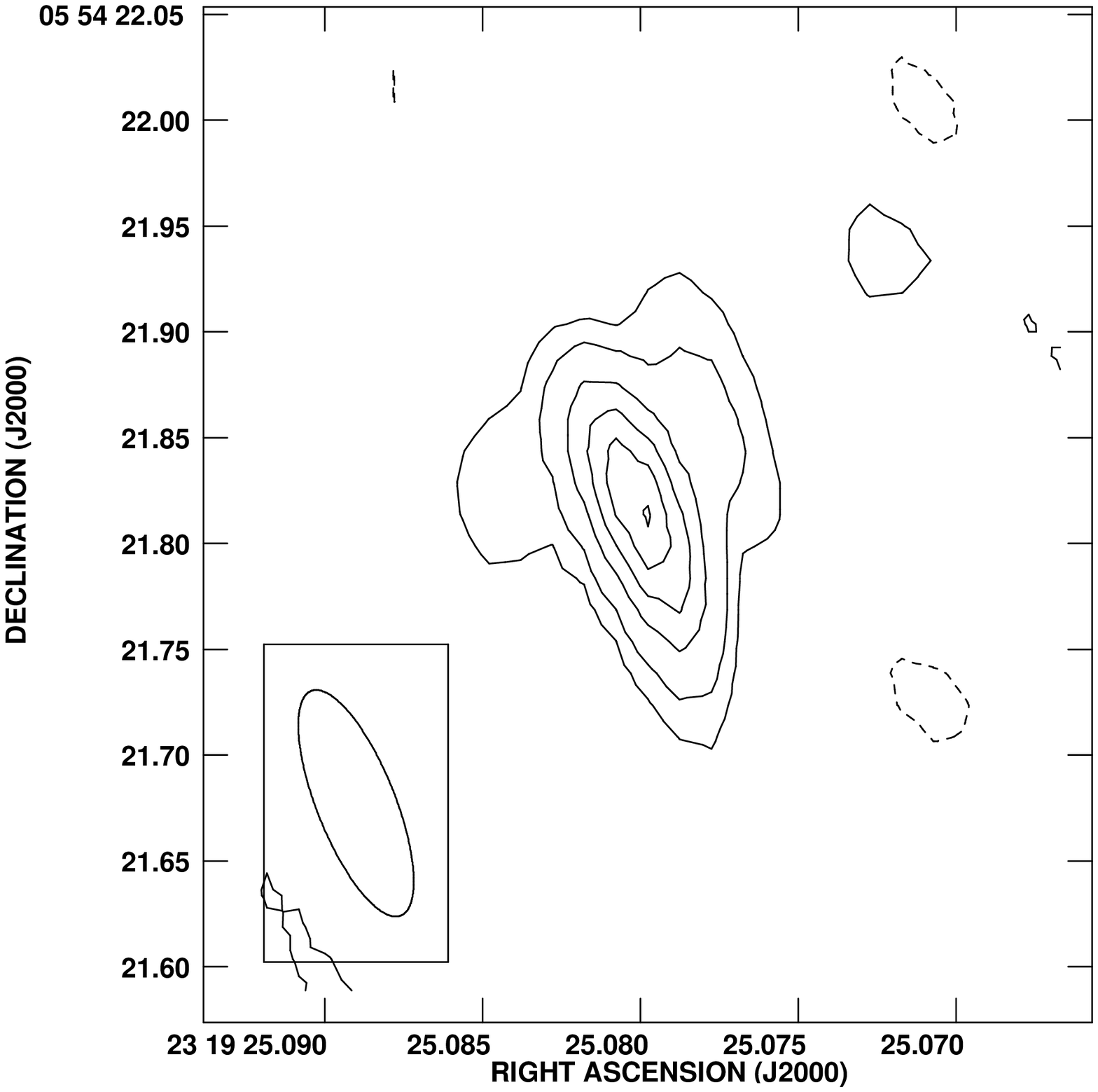}
\end{minipage}
\hfill
\begin{minipage}[t]{4.7cm}
\includegraphics[width=1.0\textwidth]{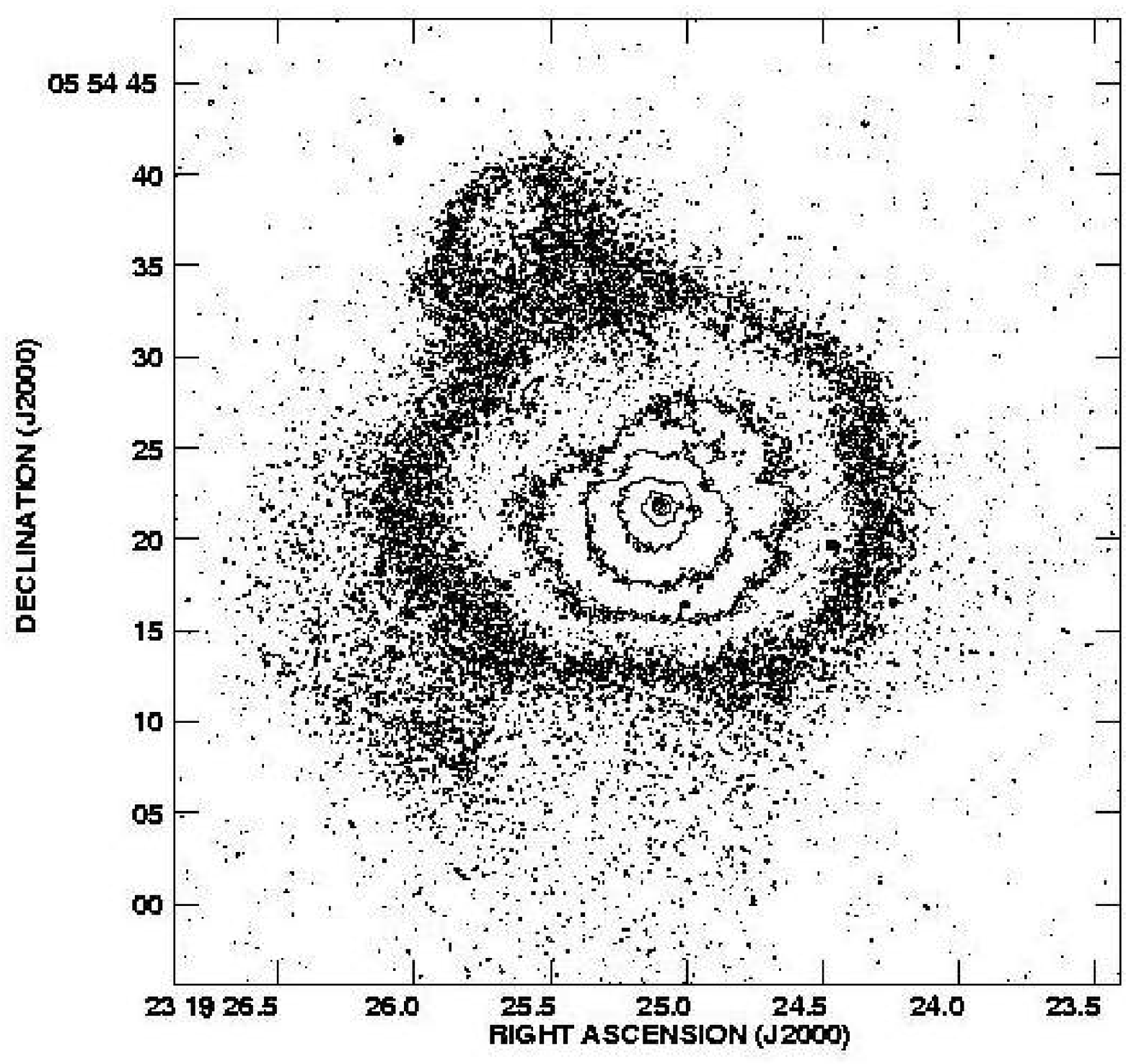}
\end{minipage}
\hfill
\hfill
\begin{minipage}[t]{4.7cm}
\includegraphics[width=1.055\textwidth]{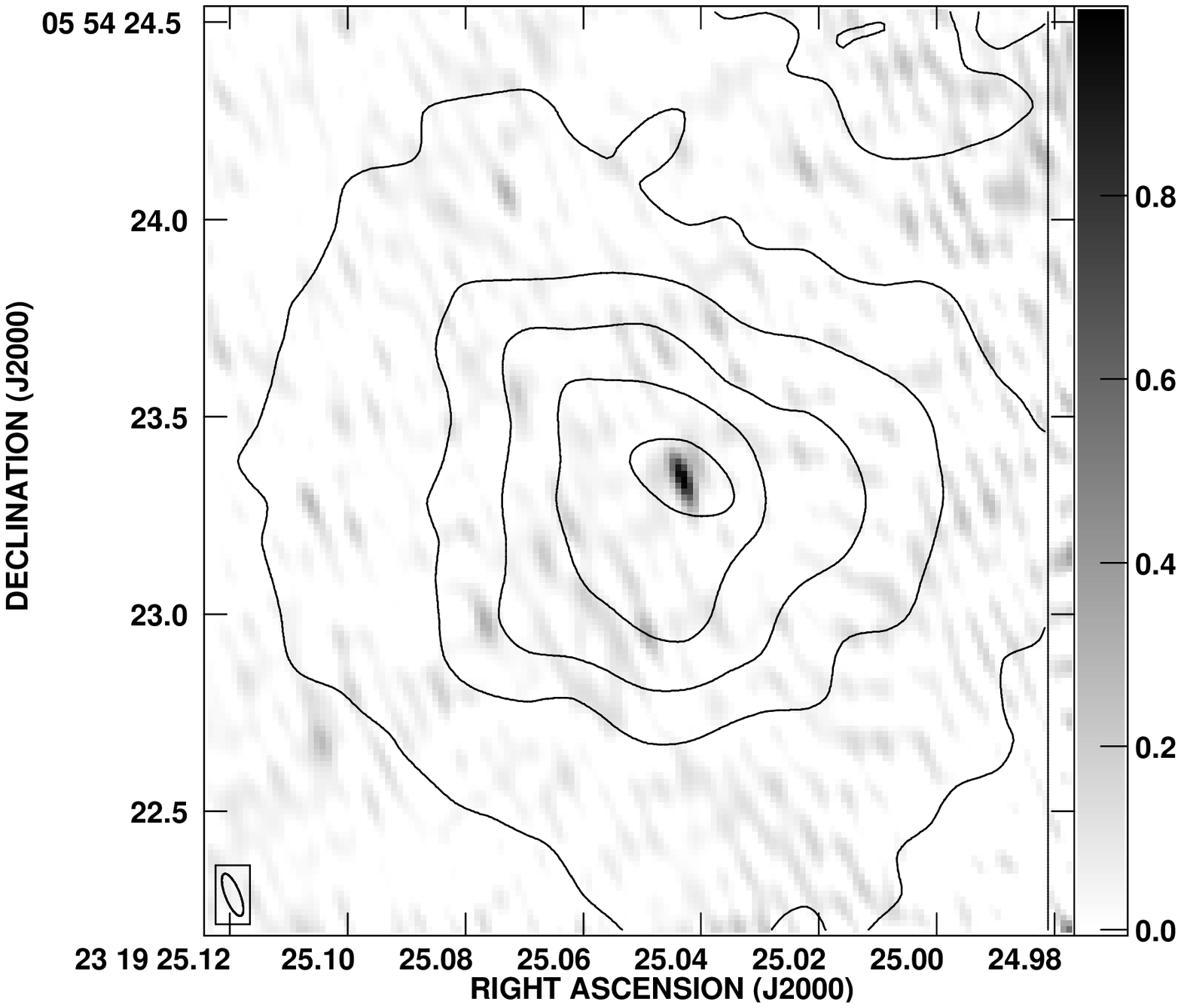}
\caption {(Left:) 5 GHz MERLIN map of IC 1481. Contour levels are
at 1.750e-4$\times$(1,2,4,8,16,32,64,128,256,512) Jy/beam. The
peak flux is 14.25 mJy/beam and the rms noise level is $\sim$ 65
$\mu$Jy/beam. (Middle:) The red continuum HST WFPC2 image of the
host galaxy. (Right:) The H$\alpha$ HST WFPC2 contour map overlaid
on 5 GHz MERLIN greyscale image.} \label{ic1481}
\end{minipage}
\hfill
\end{figure}

\vspace{0.15cm} \noindent {\bf IC 1481:} The radio map of this
LINER water maser galaxy and a possible site of ongoing merger,
shows elongated structure (Fig.~\ref{ic1481}) which agrees with
VLBI observations that have shown that water maser emission is
associated with thin, parsec-scale molecular disks located near
the spout of radio continuum jets.

\section{Discussion}

It is not widely appreciated that Seyferts, in common with radio
galaxies, may show evidence for collimated ejection and/or radio
jets. The conical or biconical morphology of the UV continuum in
Seyferts has been found to be aligned with the radio axis, and we
can postulate that the radio jets define the axis of the UV cone
and also the central engine. In this area, high resolution images
obtained by MERLIN at 5 GHz have provided the clearest evidence
for linear radio jets in Seyfert galaxies that have been
previously poorly resolved in VLA images. For this reason we
observed with MERLIN at 5 GHz and 0\arcsecpoint04 resolution, 7
Seyfert galaxies that were found to show hints of such structures
in previous lower resolution radio observations. The higher
resolution observations allow us to identify the orientation and
geometry of the nuclear disk and further examine, for the ones
that was possible, the relation of the nuclear radio structure to
the inner and large scale morphology, emission and dust lanes seen
in optical images. This way we are able to build up a more
complete picture of the physics of the low luminosity AGN.




\bibliographystyle{aipproc}   

\bibliography{SQ_proceedings}

%


%
%
%

\end{document}